\documentclass[aps,prl,preprint,groupedaddress]{revtex4-2}

\usepackage{newtxtext,newtxmath}
\usepackage{caption}

\usepackage{txfonts}

\usepackage[pdftex]{graphicx}

\begin{document}

\title{Gallium-nitride-based interference-filter-stabilized external cavity diode laser with a surface-activated-bonded output coupler}
\author{Hisashi Ogawa\thanks{E-mail: hisashi.ogawa@nichia.co.jp}, Tatsuya Kemmochi, and Tetsushi Takano}
\affiliation{R\&D Division, Nichia Corp., Anan, Tokushima 774-8601, Japan}

\begin{abstract}
We report on an interference-filter-stabilized external cavity diode laser using a gallium-nitride-based violet laser diode. 
Surface-activated-bonded glass substrates were employed as cat's eye output couplers in order to suppress power degradation due to optical damage.
From the results of a long-term frequency-stabilization test, mode-hop-free operation for about one week was demonstrated.
After a continuous operation of about three weeks, the power degradation was less than 10\%.
The results show the potential of such interference-filter-stabilized external cavity diode lasers for practical and portable quantum technologies such as atomic clocks or atomic interferometers.

\end{abstract}

\maketitle

\section{Introduction}
External cavity diode lasers (ECDLs) are used in a wide range of applications because of their features, e.g., narrow linewidth, broad tunability, compact size, and low cost.
In the field of quantum technologies using ultracold atoms or ions, in particular, ECDLs are critical light sources for laser cooling \cite{Sesko1988, Kielpinski2006} or ultra-precise spectroscopy \cite{Ludlow2007, Li2008}.
Improvements of ECDLs have contributed greatly to breakthroughs in this field \cite{Chu1985,Aspect1986,Wieman1991,Takamoto2005,Georgescu2014,Friis2018}.

There are two important directions for improving ECDLs.
The first is to extend the range of operating wavelengths.
Ultracold atoms were initially investigated mainly with alkali atoms \cite{Phillips1998a}, and gallium arsenide (GaAs)-based ECDLs that emit near-infrared light in resonance with the dipole transitions of alkali atoms \cite{Liu1981} were employed for those studies.
As research based on alkaline-earth atoms advanced, gallium nitride (GaN)-based ECDLs have been attracting attention as compact visible single-frequency light sources for cooling alkaline-earth atoms \cite{Daley2011a, Hayasaka2000}.
More recently, GaN-based ECDLs have also been applied to cooling lanthanoid atoms \cite{Golovizin2019a}.

The second important direction for improving ECDLs is to make them more compact and more wavelength-stable in order to realize more practical and portable quantum technologies \cite{Grotti2018a, Takamoto2020a, Pogorelov2021a}.
From this perspective, various types of ECDLs \cite{Numata2010a,Rauch2015,Kurbis2020} have been proposed and demonstrated in addition to the conventional diffraction grating ECDLs (G-ECDLs) \cite{Hawthorn2001,Stry2006}.
In particular, interference filter-type ECDLs (IF-ECDLs) \cite{Baillard2006} have excellent compactness and stability, since they do not need any movable elements and their cat's eye configurations make the wavelength insensitive to the angular drift of mirrors.
However, so far there are no reports of an IF-ECDL using a GaN-based laser diode (LD).

One of the difficulties in making a GaN-based IF-ECDL is the degradation of the output coupler, e.g., oxidation of the facet, deposition of carbon or $\mathrm{SiO}_2$, due to the visible light tightly focused on the output coupler surface \cite{Wang2021}.
Such degradation is more likely to occur at shorter wavelengths or higher powers, and results in power degradation and beam distortion.
Similar phenomena are observed on the output facet of GaN-based LDs, which are usually avoided by sealing the LDs in an appropriate atmosphere \cite{Wang2021, Schoedl2005}.

Here, we report on a violet GaN-based IF-ECDL in which degradation is suppressed by using two bonded glass substrates as the cat's eye output coupler.
As shown in Fig.~1, the reflection surface of the output coupler, i.e., the light-concentrating point, is at the bonding surface.
Each glass substrate acts as an end cap that decreases the light density at the entrance and exit surfaces of the output coupler.
We adopted the surface-activated bonding (SAB) method since it needs no bonding layer and suppresses voids at the bonding surface \cite{Ichikawa2016,Takagi2006,Suga1989}.

First, we conducted lifetime tests of ECDLs with different types of output couplers, and confirmed that surface-activated-bonded output couplers (SABOCs) suppress the power degradation of the ECDLs better than conventional single-plate output couplers (OCs) do.
Subsequently, two long-term wavelength-stabilization tests on the IF-ECDL with SABOC were performed.
Its wavelength was continuously stabilized to 410.7~nm, the resonant wavelength of thulium atoms.
During the second test of three-weeks continuous operation, the power degradation remained at less than 10\%.
Furthermore, one-week maintenance-free operation was achieved.
The results show that our method is suitable for light sources for atomic clocks or atom interferometers using alkaline earth or lanthanoid atoms \cite{Takamoto2020a, Hu2017, Golovizin2019a}.

\section{Design of the IF-ECDL}

\begin{figure}
\centering
\includegraphics[width=0.99\textwidth]{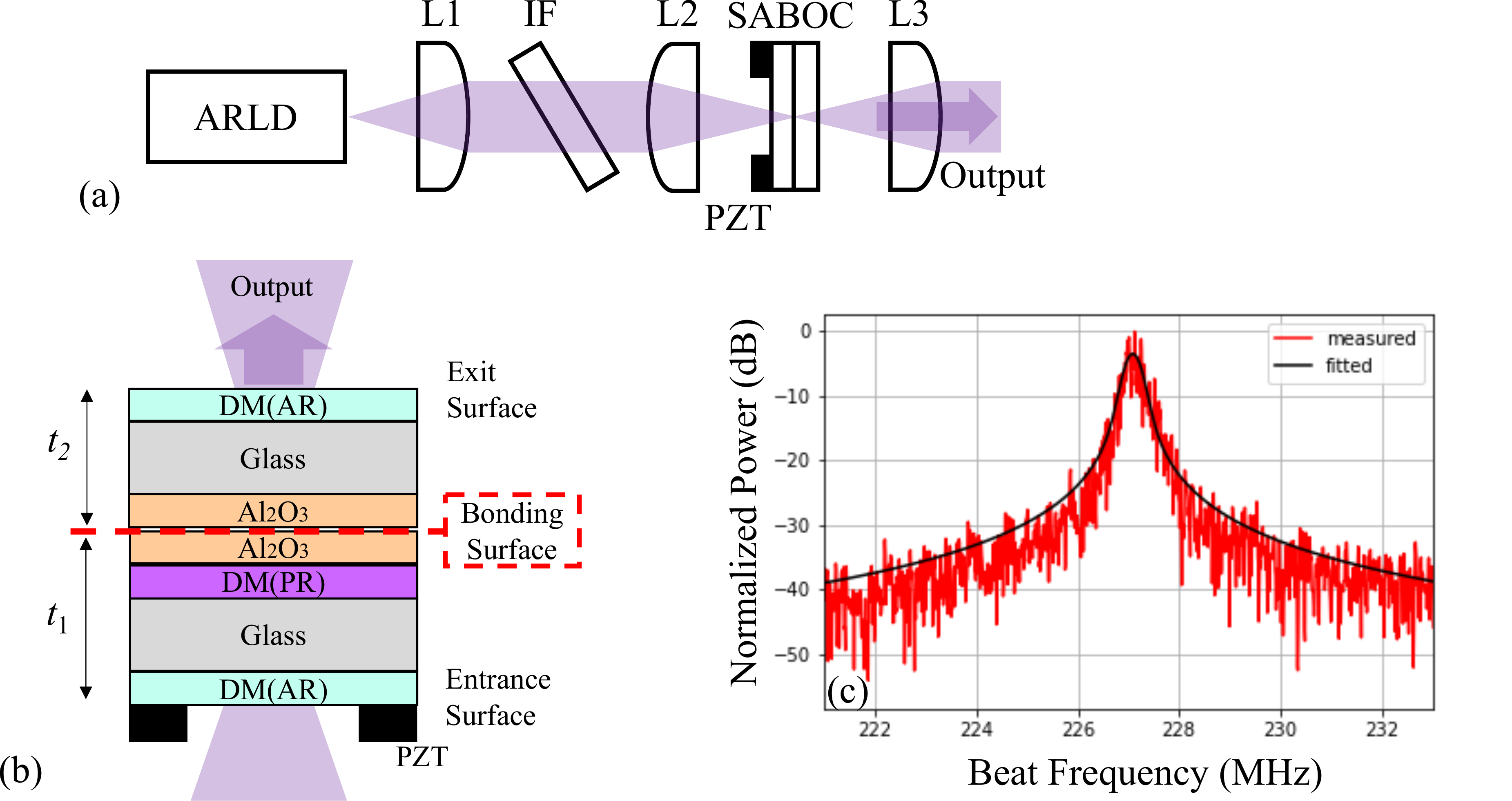}
\caption{(a) Schematic diagram of the IF-ECDL with SABOC. ARLD: anti-reflection coated LD, L1: collimation lens, IF: interference filter, L2 and L3: lenses for making cat's eye configuration, SABOC: surface-activated-bonded output coupler, PZT: piezoelectric transducer. (b) Overview of SABOC. DM: dielectric mirror coating, AR: anti-reflection, PR: partial-reflection. (c) Beat-note spectrum of the two free-running IF-ECDLs.}
\end{figure}

Figure~1(a) shows a schematic diagram of the IF-ECDL with a SABOC.
Light emitted from an anti-reflection (AR)-coated GaN-based LD (Nichia NDVA416) was collimated by a lens L1 (Thorlabs C671TME-A ($f$ = 4.02~mm)), transmitted through an interference filter IF (LaserOptik L-17182-01), and entered a cat's eye part.
The interference filter had a transmission linewidth of about 0.2~nm and a peak transmission efficiency of about 70\%.
The cat's eye part consisted of an OC with a piezoelectric transducer (PZT) placed between lenses L2 and L3 (L2 and L3 were A375TM-A ($f$ = 7.5~mm) for comparison of OCs (see Section~3)  and C280TMD-A ($f$ = 18.4~mm) for long-term frequency stabilization tests (see Section~4)).
The LD and the optical elements were installed in an aluminum housing, and the temperature of the housing was controlled with a Peltier module.
The length of the cavity was about 50~mm (about 100~mm round trip), and thus, the free-spectral range was about 3~GHz.
The mode-hopping free scanning range was around 1~GHz when the cavity length was scanned with the PZT.

An overview of the SABOC is shown in Fig.~1(b).
In SAB, the bonding surfaces were sputter-etched and activated using argon fast ion beam, and subsequently bonded at room temperature under vacuum.
The base pressure of the vacuum chamber was less than $7.0\times10^{-6}$~Pa.
Special care was taken when polishing the glass substrates with chemical mechanical polishing (CMP) to ensure atomically flat surfaces.
The entrance and the exit surfaces of the SABOC were AR coated whereas the bonded surface has a dielectric multi-layer partial-reflection (PR) coating.
The bonding strength of the SAB method is dependent on the materials involved \cite{Ichikawa2016,Takagi2006,Suga1989}.
In order to ensure the bonding strength, we chose $\mathrm{Al}_2\mathrm{O}_3$ as the top coating layers at the bonding surface.
The reflectivity of the PR coating, including $\mathrm{Al}_2\mathrm{O}_3$ layers, was about 50\% at a wavelength of 410~nm.
It is reasonable to assume that there was no light absorption at the bonded interface because no adhesive material was used.
The thicknesses of the glass substrates on the incident side and on the output side are denoted as $t_1$ and $t_2$, respectively.
Assuming that the $e^{-2}$ beam waist on the reflective surface is about 4~$\mathrm{\mu m}$, the light density on the entrance and the exit surfaces are decreased by about 1/80 for $t_{1,2}$ = 1~mm and about 1/320 for $t_{1,2}$ = 2~mm, which is expected to suppress mirror degradation and prevent power degradation of the IF-ECDL.
We tested both $t_{1,2}$ = 1 and 2~mm for comparison of OCs (see Section~3), and adopted $t_{1,2}$ = 2~mm for long-term frequency stabilization tests (see Section~4).

In order to estimate the laser linewidth, we built two IF-ECDLs of identical design and measured a beat-note spectrum of the two lasers.
Figure~1(c) shows the beat-note spectrum when the IF-ECDLs are free-running (without a frequency stabilization).
From the Voigt fitting of the spectrum, the linewidth of each laser was estimated to be about 300~kHz FWHM, which was dominated by Gaussian component \cite{Domenico2010}.
Note that the resolution bandwidth of the spectrum analyzer was 10~kHz and the sweep time of the single frame was about 1.6~second.

\section{Comparison of output couplers}

\begin{figure}
  \centering
  \includegraphics[width=0.99\textwidth]{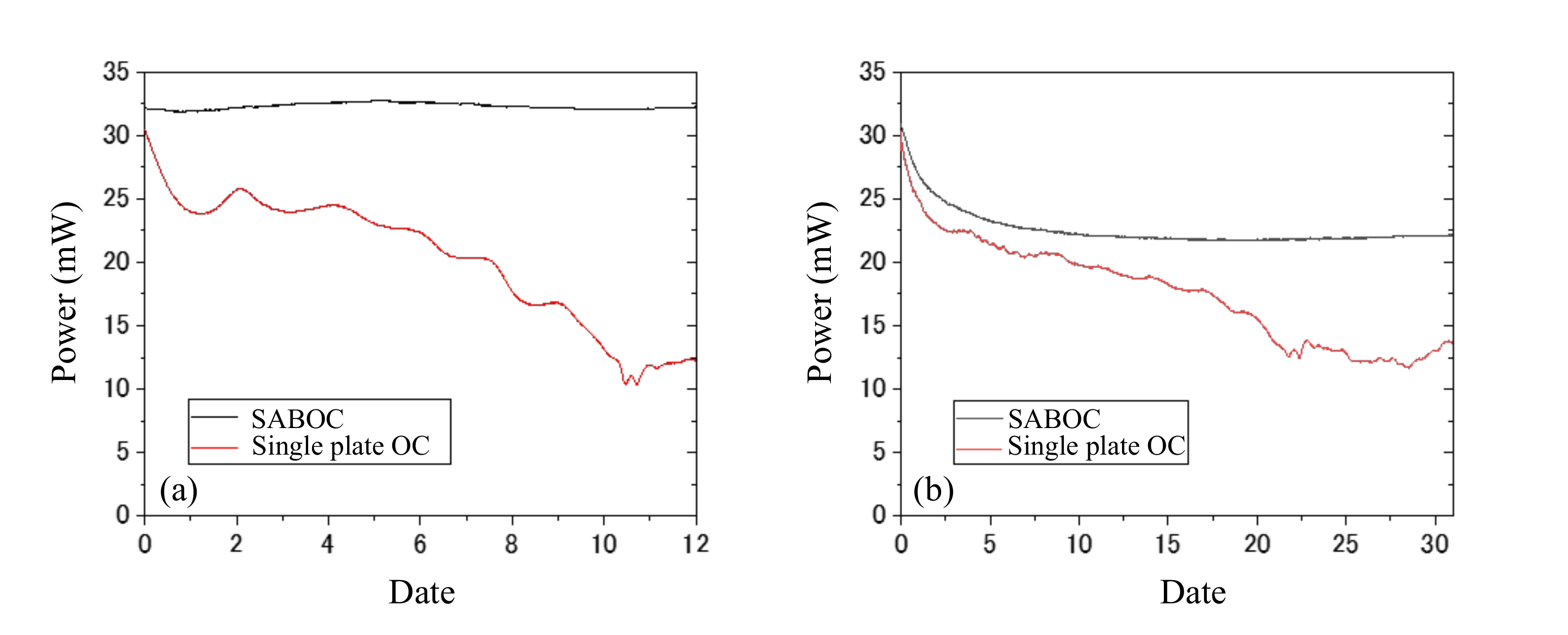}
  \caption{Comparison of power changes of ECDLs during accelerated life tests. Black curves: with SABOC, red curves: with single-plate OC. (a) Result with $t_1$ = $t_2$ = 1~mm. (b) Result with $t_1$ = $t_2$ = 2~mm.}
  \label{Fig2}
\end{figure}

To confirm the effectiveness of the SABOC, we compared the powers of ECDLs with a SABOC and with a conventional single-plate OC.
The powers of both ECDLs were set to 30~mW at the start of the measurement, and the change in power was recorded.
By removing IFs and using the lenses L2 and L3 with a shorter focal length ($f$ = 7.5~mm), the ECDLs used here had higher intra-cavity power and more tightly focused light spots on the reflection surfaces than the IF-ECDL used in Section 4, and thus, the comparison was relatively accelerated.
The lasers oscillated in multiple longitudinal modes ranging 409~nm $\pm$ 1~nm.
During the measurements, the ECDL with the SABOC and the ECDL with the conventional single-plate OC were placed next to each other in order to share the conditions, such as air temperature and cleanliness.

Figure~2(a) shows the result with $t_1$ = $t_2$ = 1~mm, whereas Fig.~2(b) shows the result with $t_1$ = $t_2$ = 2~mm.
In both cases, the power of the ECDL with the single-plate OC degraded faster than the other one with the SABOC.
The power degradations of the ECDLs with the single-plate OCs were accompanied by unstable kinks.
Such power degradation with kinks has been observed in previous studies on facet degradation of unsealed LDs \cite{Wang2021}.
In Fig.~2(a), the power degradation of the ECDL with the SABOC remained within few percent for 12 days.
On the other hand, the power in Fig.~2(b) decreased about 25\% in first 5 days and then was maintained for the next 25 days.
We presume that the power degradation in the first 5 days in Fig.~2(b) was caused by slight positional shifting of the optical elements, e.g., focus shifting due to positional shifting of the lenses, as the measurement started not long after the alignment was changed significantly.
We think that this causes no problem in practical use since the power was maintained after that.
In all four cases shown in Fig.~2, the power of the LD without OC was checked before and after the life test, and the degradation was less than 5\%.
Therefore, we concluded that the differences in the degradation rate were mainly due to the OCs, and that the SABOCs successfully suppressed the reduction in power.

\section{Long-term frequency stabilization tests}

Two long-term continuous-operation tests with frequency stabilization were performed.
The IF was installed to operate the laser in a single longitudinal mode.
The focal length of the lenses L2 and L3 were $f$ = 18.4~mm and the thickness of OC was $t_1$ = $t_2$ = 2~mm.
A wavemeter stabilized the the IF-ECDL to a wavelength of about 410.7~nm, which is nearly resonant with the cooling transition of thulium atoms \cite{Golovizin2019a}, by tuning the voltage applied to the PZT.
Data acquisition period of the wavemeter was about 50~ms.
The change of 1 in the servo signal corresponded to the cavity length shift of approximately 250~nm (about 2~pm compensation in the resonant wavelength).
The results are shown in Fig.~3.

\begin{figure}
  \centering
  \includegraphics[width=0.9\textwidth]{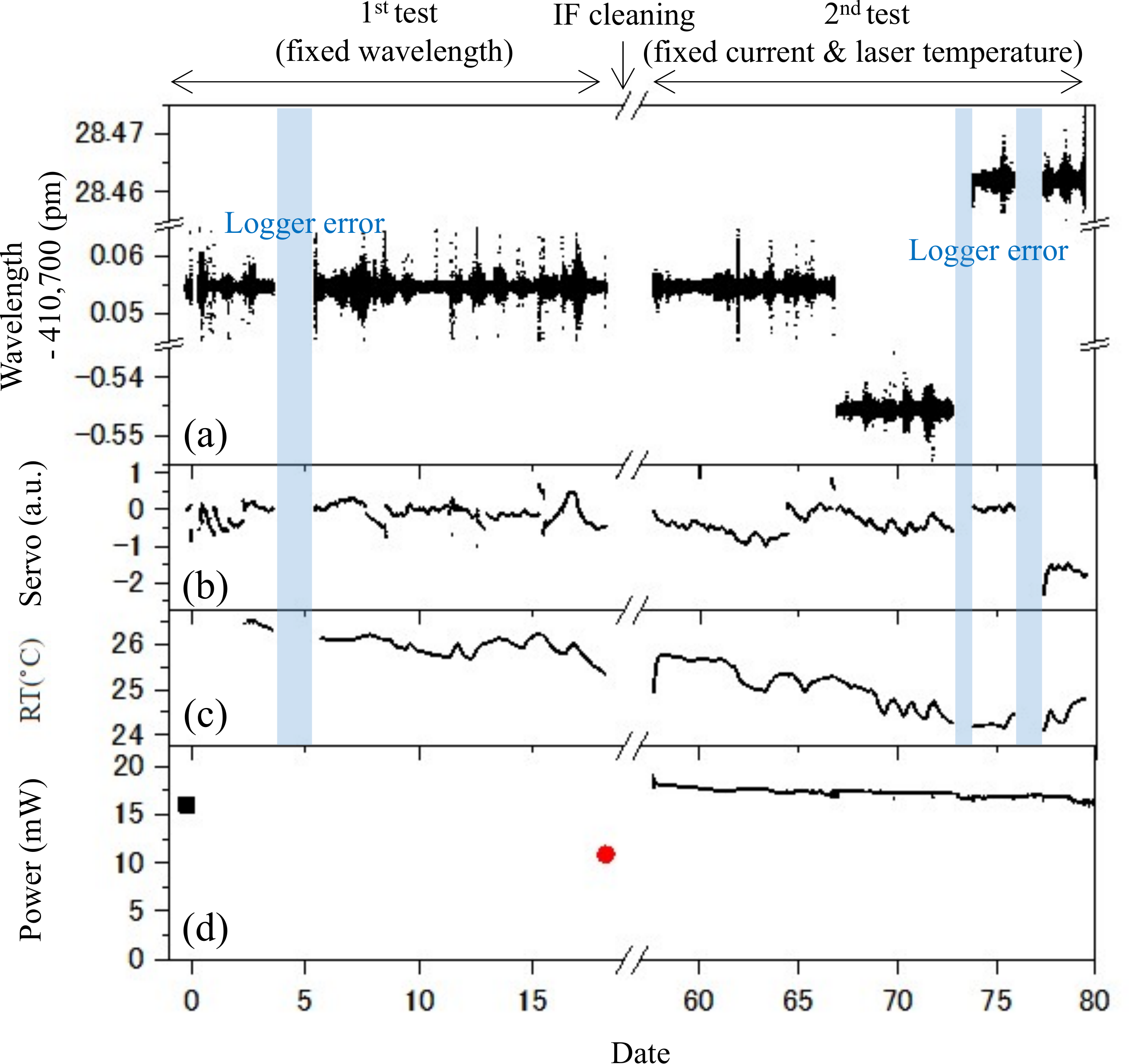}
  \caption{Results of the long-term frequency stabilization tests. (a) Behavior of the wavelength. (b) Feedback signals of piezoelectric servo controller. (c) Room Temperature. (d) Power of the IF-ECDL.}
\end{figure}

\begin{figure}
  \centering
  \includegraphics[width=0.7\textwidth]{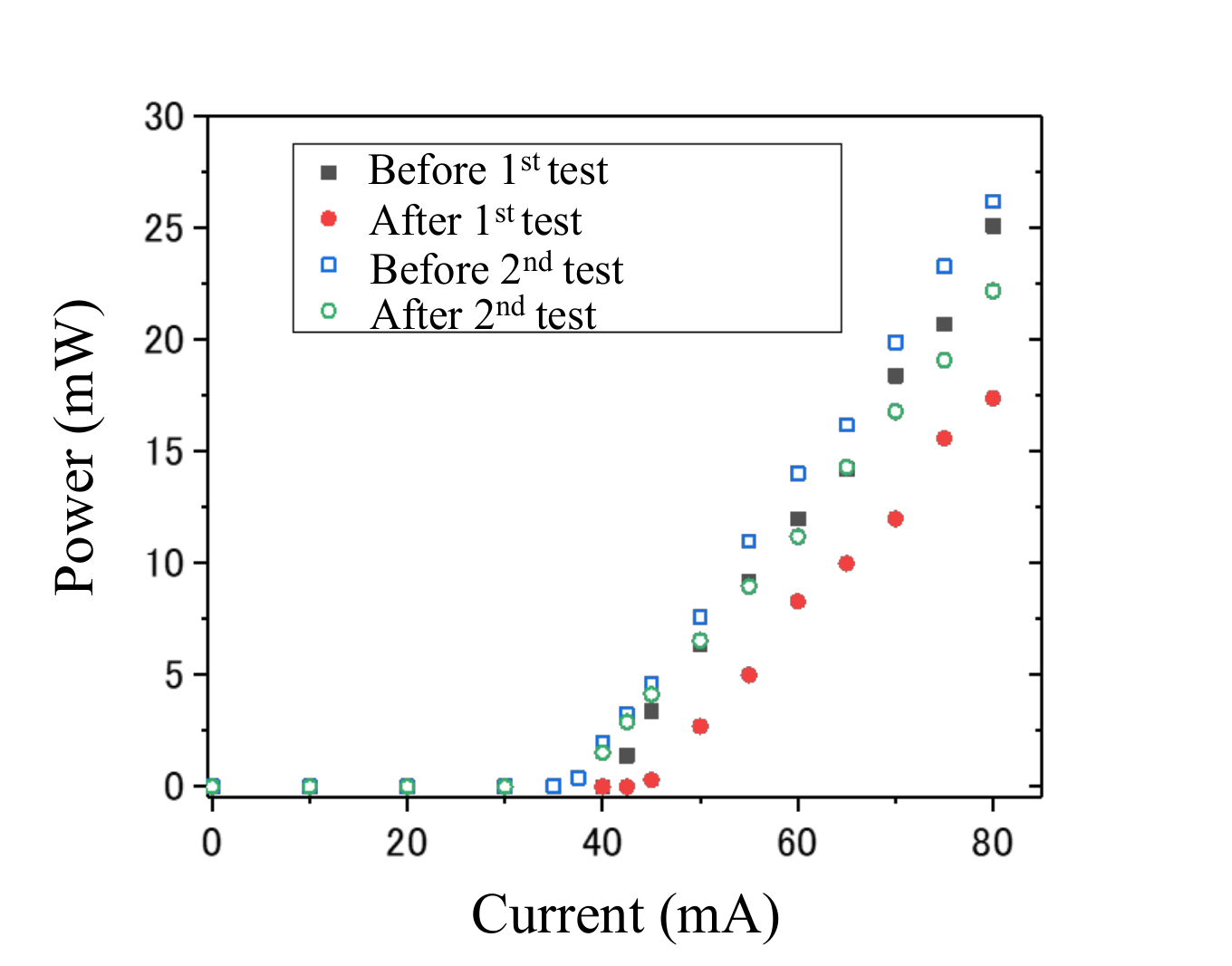}
  \caption{Power--current characteristics of the IF-ECDL before and after each tests.}
\end{figure}

Figure~3(a) shows the behavior of the wavelength.
In the first test, mode hopping occurred about once a day, as shown by the servo signal's jumps in Fig.~3(b).
When mode hopping occurred, the laser diode current or temperature of the ECDL housing was manually adjusted to bring the wavelength back.
The time used for the manual relocking was short compared to the time scale of the graph, and thus, the wavelength values in Fig.~3(a) appear to be plotted almost continuously.
After the 18-day first test, the laser power decreased by about 30\%, which is shown by the black square and red dot in Fig.~3(d).
The cause of the power reduction was identified as contamination of the IF surface by spillage of poorly cured adhesive which was used for bonding the IF to a holder.
The IF was cleaned before the 2nd test, and the power was recovered as shown in Fig.~4.

Since the mode hopping rate decreased after cleaning the IF, we conducted the 2nd test only with PZT feedback, without manually adjusting the laser current and temperature.
Even in the maintenance-free operation, the wavelength remained locked for about one week, from the date 57.7 to 64.4.
Around the 67th day, a mode-hop occurred again in response to the change in the room temperature.
The experiment was continued by changing the locking wavelength while keeping the laser current and temperature unchanged.
The output power variation was within 10\% during the 2nd test of about three-weeks continuous operation.

Since uptime of the stabilization is important in some applications of optical lattice clocks \cite{Takano2016, Kobayashi2020}, the uptime during the test was also calculated to demonstrate applicability of the IF-ECDL to such usage.
Even with the contamination of the IF, the uptime in the 1st test was about 97.8\% excluding the duration of data missed due to a logger error which is shown by the blue region in Fig.~3.
The uptime improved to about 99.8\% in the 2nd test.
These values are sufficiently high compared to the uptime of the optical lattice clocks in previous studies (73\% in \cite{Takano2016} and 93.9\% in \cite{Kobayashi2020})

Figure~4 summarizes the change in the laser power before and after the tests.
Comparing before and after the 1st test, the lasing threshold increased and the power decreased by about 30\% at a driving current of 80~mA, which was caused by contamination of the IF by poorly cured adhesive as already discussed.
On the other hand, in the case of the 2nd test, the lasing threshold did not change much, and the decrease in power was less than 10\%.
Therefore, we conclude that the degradation of the LD and the SABOC was small enough to keep the output power variation within 10\% during the three weeks continuous operation.
Note that the total time of use of the LD and the SABOC in the two tests was 39 days.
We consider that longer operation with less mode-hopping is possible if mechanical, temperature, or barometric stability of the system can be improved by, for example, precise positioning using laser welding \cite{Ohmae2021} or dual temperature control \cite{Takamizawa2016a}.

\section{Summary and future prospects}

We developed a 410~nm GaN-based IF-ECDL with a SABOC, which can be used for thulium laser cooling.
We successfully observed an improvement in lifetime by employing the SABOC, and performed long-term frequency stabilization tests to show the practicality of our approach.
As a result, maintenance-free operation for about one week was achieved, and the power degradation after three weeks of continuous operation was less than 10\%.
In the future, precise positioning by laser welding \cite{Ohmae2021} or dual temperature control might enable longer maintenance-free operation \cite{Takamizawa2016a}.
These results suggest that our approach could be a pivotal technology to improve the portability and long-term operation of atomic clocks or atomic interferometers using alkaline earth and lanthanide atoms \cite{Takamoto2020a,Hu2017,Golovizin2019a}.

\bibliography{Collection}

\end{document}